\documentclass{article}

\usepackage{amsfonts,amsmath,amsthm,parskip}
\usepackage{graphicx}
\usepackage{subcaption}
\usepackage{booktabs}
\usepackage{xcolor}
\usepackage[alpine]{ifsym} 
\usepackage{url}
\usepackage[square,numbers]{natbib}

\usepackage[sc,osf]{mathpazo}   
\usepackage[euler-digits,small]{eulervm}
\usepackage{upgreek}
\usepackage{authblk}

\usepackage{geometry}

\title{Comment on ``Which is greater: $e^{\pi}$ or $\pi^{e}$? An unorthodox physical solution to a classic puzzle''}

\author[1]{Roderick M.\ Macrae\thanks{ORCiD: 0000-0002-9083-2518}}

\affil[1]{Department of Chemistry, Marian University, Indianapolis, Indiana 46222-1997}

\begin{document}

\maketitle

In a recent Note\cite{vallejo_which_2024}, Vallejo and Bove provide a physical argument based nominally on the second law of thermodynamics as a way of resolving the  mathematical question appearing in the title. 
A remarkable aspect of their argument is that it does not depend on the numerical value of $\pi$, 
because $e^{x} \ge x^{e}$ for all positive $x$,
with equality occurring only when $x = e$. Moreover, their argument does not depend on the validity of 
the second law but is rather a limited proof of it for this particular case.

Their  argument is based on a scenario in which an incompressible solid body $A$ with constant heat 
capacity $C$  at initial temperature $T_{1} = \pi$ in the units of some absolute temperature scale 
is placed in contact with an ideal reservoir $B$ at initial temperature $T_{2} = e$ in the same units.  
The system evolves irreversibly to equilibrium at  the temperature of the 
reservoir. In these units $\Delta S_{A} = C (1 - \ln \pi)$ and $\Delta S_{B} = C\left(\pi/e - 1\right)$,
 leading to an overall entropy change $\Delta S = C \left(\pi/e - \ln \pi \right)$. 
 Invoking the second law $\Delta S > 0$ for an irreversible process, the authors obtain $\pi > \ln \pi^{e}$ and thus $e^{\pi} \ge \pi^{e} $.

The argument appears to depend on the value of $e$ through $\ln e = 1$ as one of the steps in 
obtaining $\Delta S_{A}$ 
(the more general case is discussed below),
but does  not make use of the numerical value of $\pi$, for example by determining the direction 
of heat flow through its relation with $e$. Thus, the argument  and consequently the result must be independent of the value of $\pi$ 
provided it is real and positive. This comment investigates this situation more fully.

Stepping back for a moment from the authors' specific choice of temperature scale but following the same 
arguments, the more general expression for the overall entropy change can be shown to be
\begin{equation}
\label{eqn:eq1}
\Delta S  = C \left( \ln T_{2} - \ln T_{1}  + \frac{T_{1}}{T_{2}} - 1\right),
\end{equation} 
where the first two terms represent $\Delta S_{A}$ and the last two $\Delta S_{B}$. 
Letting $T_{2}/T_{1}$ equal $x > 0$ (both are absolute temperatures), we may rewrite this as
\begin{equation}
\label{eqn:eq2}
\Delta S  = C \left( \ln x + \frac{1}{x} - 1\right),
\end{equation} 
where the 
variable part 
can easily be shown to have a single minimum of zero occurring at $x = 1$, and a positive second derivative. 
This can be interpreted to imply that 
{\em heat flow between two bodies at different temperatures is always accompanied by an increase 
in total entropy}, and is, in slightly different form, an argument commonly found  in thermodynamics 
 for entropy increase as an indicator of the direction of spontaneous change\cite{mcquarrie1997physical}. 
 In this particular case the entropy change is positive quite independent of the direction of heat flow,
 that is,  of whether the reservoir is hotter or cooler than the body.

The choice by the authors of the rather special unit system in which $T_{2} = e$ allows eqn.\ \ref{eqn:eq1} to be rewritten in the form 
\begin{equation}
\label{eqn:eq3}
\Delta S  = C \left( \frac{T_{1}}{e} - \ln T_{1}  \right),
\end{equation} 
because the second term in  eqn. \ref{eqn:eq1}, which equals 1 in these units, 
cancels with the fourth term, which is equal to 1 in any 
units. As a restriction of eqn. \ref{eqn:eq1} to a particular unit scale, 
eqn. \ref{eqn:eq3} inherits its qualitative behavior and is $\ge 0$ for all values of $T_{1}$, 
whether greater than or less than $e$. The case $T_{1} = \pi$ is a rather arbitrary special case. 
Vallejo and Bove note the more general inequality in their equation 7.

\begin{figure}[h]
 \centering
  \includegraphics[width=10cm]{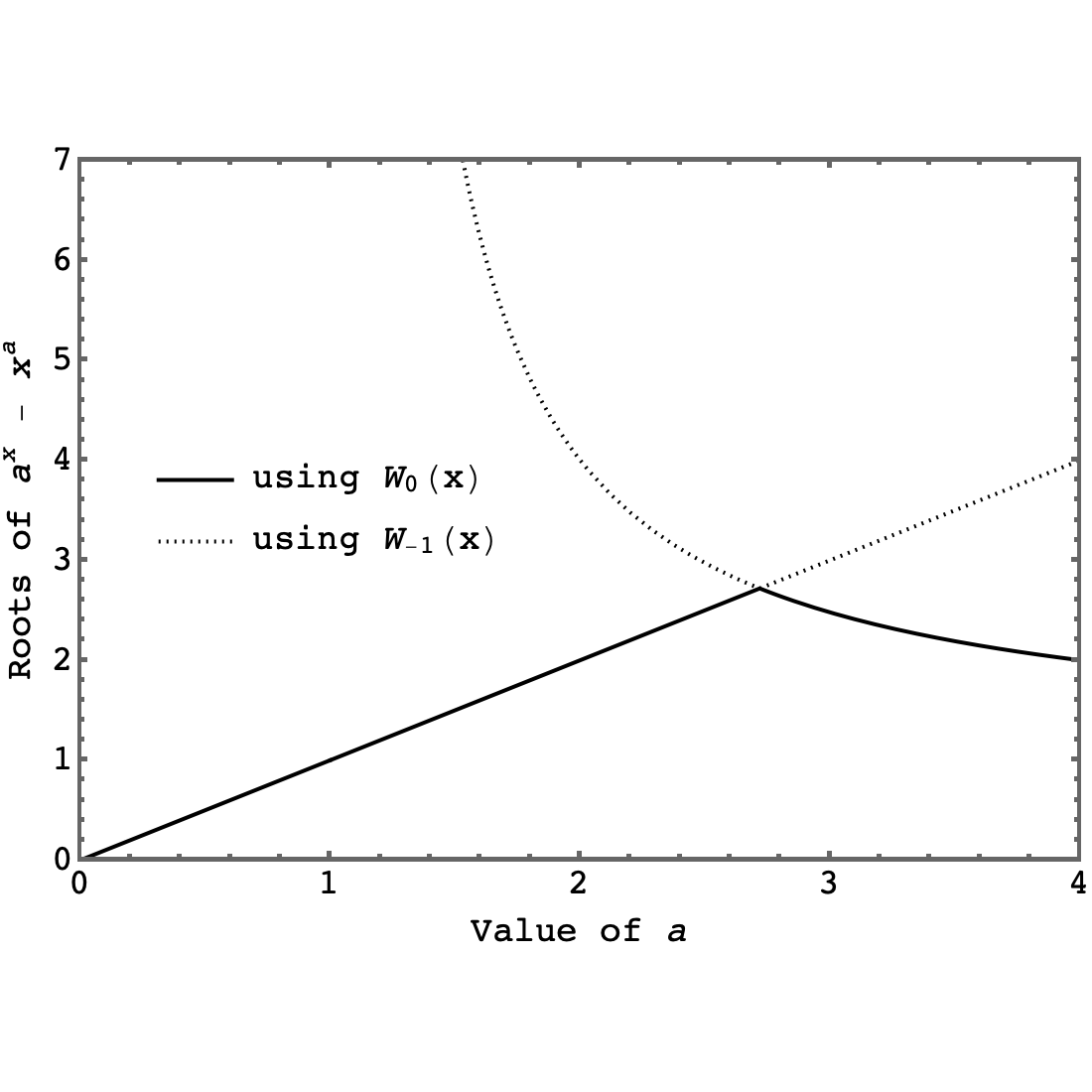}
   \caption{Roots of $f(x) = a^{x} - x^{a}$ showing the branches originating from $W_{0}(x)$ and $W_{-1}(x)$. The branches coincide at $a = e$.}
\label{fig:roots}
  \end{figure}

While the above arguments show that the numerical value of $\pi$ is not important in this case, the numerical 
value of 
Euler's number 
$e$ is critical, as the function $f(x) = a^{x} - x^{a}$ is uniformly non-negative over $(0,\infty)$ \emph{only} 
for $a = e$. 
For $a > 1$, $f(x)$ has in general two roots and is negative between them, while for $0 < a \le 1$ there is only one root, lying in the same interval, and approaching zero from above as $a \to 0$. 
The solutions to $f(x) = 0$ take the form 
\begin{equation}
x = -a W\left(-\ln(a)/a\right)/\ln (a), 
\end{equation}

where $W$ is the Lambert $W$-function. The behavior of the roots of $f(x)$ is a consequence of the two-valued nature of $W(x)$ over $-1/e < x < 0$.
In this region the branch with values greater than -1 is known as the principal branch and typically denoted $W_{0}(x)$, while the secondary (lower) branch is usually denoted  $W_{-1}(x)$\cite{weisstein_lambert_nodate,corless_lambertw_1996}. Figure \ref{fig:roots} shows the dependence on $a$ of the roots of $f(x)$ in the vicinity of $a = e$.

Following on from the arguments of Vallejo and Bove, equation \ref{eqn:eq2}, when coupled with the second law, 
might be taken as a ``proof'' that $1 - 1/x$ is a lower bound for $\ln(x)$, as is well known.  
%
%
However, the inequality follows straightforwardly from the analytical properties of the function, 
and the second law need never be invoked. Rather, the inequality 
acts as a demonstration that the second law is valid for this model system.


Interest in thermodynamic ``proofs'' of mathematical inequalities appears to have begun with Landsberg's short, citation-free article applying the first and second laws to $n$ identical heat reservoirs initially at different temperatures to affirm the inequality 
between the arithmetic and geometric means \cite{LANDSBERG19781}. 
As noted in a brief historical article by Deakin\cite{deakin1999}, however, the argument  dates 
back to P.\ G.\ Tait in 1868\cite{Tait1868}, and was used as an exercise in Sommerfeld's book on 
thermodynamics and statistical mechanics \cite{sommerfeldthermodynamics}; by 1980 Landsberg had become aware of 
Sommerfeld's work \cite{LANDSBERG1980}. A collection by Tykodi of similar inequalities supported by 
model systems was published in this journal in 1996\cite{tykodi1996}, 
and 
a demonstration by Plastino {\em et al.} of thermodynamic support for Jensen's inequality,
of which the inequality of the arithmetic and geometric means is a consequence, 
was published the following year\cite{plastino1997}. Over time, the framing of these examples has shifted, noting that they are not strictly ``proofs'' \cite{deakin1999}, and are more correctly characterized as demonstrating mathematical inequalities\cite{tykodi1996}. 

A recent article in  this 
journal by Johal\cite{johal2023} returns to the source from which Tait built his original observation, namely 
a paper
by William Thomson (Lord Kelvin)
on the extraction of all available work from an unequally heated space by means of a heat engine\cite{Thomson1853}.
Tait updated and discretized Thomson's result to determine that for a set of identical masses, the final temperature
after such a process is the geometric mean of their initial temperatures,
while the temperature achieved by thermal equilibration is the higher arithmetic mean\cite{Tait1868}. 
Limiting consideration to two masses for simplicity, Johal notes that a more edifying interpretation of the thermalization process can be obtained by dividing it into two steps: a reversible one in which all available work is extracted until the bodies are at the same temperature (the geometric mean of their original temperatures), and a second one in which the same quantity of energy is returned as heat and the bodies are warmed to the arithmetic mean of their original temperatures. 
The input of heat in this second step is a useful pedagogical illustration that the final entropy of the system 
must be higher in accordance with the second law. 
Similar arguments were made previously by Pyun\cite{Pyun1974} and Leff\cite{Leff1977}. 
All such arguments depend on the positivity of the heat capacity of material bodies, which 
may not be universally valid \cite{Roduner2023}. 
An additional point made by Leff and worth reiterating here is that although the second law in the form of entropy increase 
is demonstrated rather than assumed by Vallejo and Bove, 
it is a central requirement of their example that temperature equilibration - one of the observed macroscopic phenomena leading to the invention of the entropy concept - takes place.

\bibliography{greaterbib}{}
\bibliographystyle{unsrtnat}

\end{document}